# Modeling the Effect of Dissolved Hydrogen Sulfide on $Mg^{2+}$-water Complex on Dolomite {104} Surfaces

*Zhizhang Shen[1], Yun Liu[2,3], Philip, E. Brown[1], Izabela Szlufarska*[2,3], and Huifang Xu*[1,3]*

[1]Department of Geoscience, University of Wisconsin-Madison, 1215 W. Dayton Street, Madison, WI53706, USA

[2] Department of Materials Science and Engineering, University of Wisconsin-Madison, 1509 University Avenue, Madison, WI53706, USA

[3] Materials science program, University of Wisconsin-Madison, 1509 University Avenue, Madison, WI53706, USA

*Corresponding authors

hfxu@geology.wisc.edu

Tel: 608-265-5887  Fax: 608-262-0693

szlufarska@wisc.edu

Tel: 608-265-5878  Fax: 608-262-8353

ABSTRACT

The key kinetic barrier to dolomite formation is related to the surface $Mg^{2+}$-$H_2O$ complex, which hinders binding of surface $Mg^{2+}$ ions to the $CO_3^{2-}$ ions in solution. It has been proposed that this reaction can be catalyzed by dissolved hydrogen sulfide. To characterize the role of dissolved hydrogen sulfide in the dehydration of surface $Mg^{2+}$ ions, *ab initio* simulations based on density functional theory (DFT) were carried out to study the thermodynamics of competitive adsorption of hydrogen sulfide and water on dolomite (104) surfaces from solution. We find that water is thermodynamically more stable on the surface with the difference in adsorption energy of -13.6 kJ/mol (in vacuum) and -12.8 kJ/mol (in aqueous solution). However, aqueous hydrogen sulfide adsorbed on the surface increases the $Mg^{2+}$-$H_2O$ distances on surrounding surface sites. Two possible mechanisms were proposed for the catalytic effects of adsorbed hydrogen sulfide on the anhydrous Ca-Mg-carbonate crystallization at low temperature.

**INTRODUCTION**

Dolomite (CaMg(CO$_3$)$_2$), which used to be ubiquitous in the geological past, is rarely found in Holocene sediments.[1] Its rare occurrence in the modern sediments defies the geological notion that the present is the key to the past. This contradiction is at the heart of the famous "dolomite problem". Attempts to synthesize dolomite inorganically under ambient environment (low temperature, <50°C) have been largely unsuccessful[2], although a recent study[3] indicates that natural dolomitization could happen at as low as 40°C. It is now generally accepted that the dolomite problem lies in the high kinetic barrier caused by the magnesium hydration at low temperature that hinders dolomite formation.[4-5]

Based on the observations that modern dolomite formations are usually associated with environments where anaerobic microorganisms, including sulfate-reducing bacteria, are active[6-12], microorganisms are believed to help overcome the kinetic barrier and to promote dolomite formation. It was proposed that sulfate reducing bacteria are related to dolomite crystallization in nature[7], and high-Mg calcite and Ca-rich disordered dolomite have been synthesized at low temperature under laboratory conditions with sulfate reducing bacteria[7,9]. In addition, a recent work showed that dissolved hydrogen sulfide, as a product of bacterial sulfate reduction, is an eligible catalyst for dolomite crystallization[13,14]. The amount of MgCO$_3$ in the precipitating Ca-Mg-carbonate increases as concentration of the dissolved sulfide increases.[14] The dissolved hydrogen sulfide may take the role of a catalyst for enhancing Mg incorporation into the structure and crystallization of dolomite.

Despite these advances, the mechanism of how dissolved hydrogen sulfide, or catalysts in general, helps overcome the kinetic barrier is uncertain. The dolomite growth involves two important steps: First, hydrated $Mg^{2+}$ ions from the solution are adsorbed onto the dolomite surface and then the surface hydrated $Mg^{2+}$ ions attract $CO_3^{2-}$ ions from the solution (Figure 1). The positive role of catalysts in the first step has been supported by the evidence that disordered dolomite has been synthesized at low temperature by using carboxymethyl cellulous (CMC) with carboxyl functional groups[15], which can dewater and complex with Mg ions forming a $[Mg(H_2O)_5(R\text{-}COO)^+]$ complex.[16,17] It was proposed that Mg-carboxyl complex may change coordination environment of Mg[16] and requires much lower energy (56.9 kJ/mol) for carbonation than $Mg(H_2O)_6^{2+}$.[18] However, $HS^-$ has been shown to neither have much effect on lowering the $Mg^{2+}$ dehydration barrier in the aqueous $Mg^{2+}$-water complex nor interact with the remaining five water molecules in the first solvation shell of $Mg^{2+}$.[19] The adsorption of magnesium onto a growing calcite crystal from the solution is energetically favorable according to a previous simulation study.[20] Lippmann also argues that there is no significant energy barrier for $Mg^{2+}$ to be adsorbed onto the surface, and there is no need for $Mg^{2+}$ to be entirely dehydrated in order for it to be adsorbed.[4] The author further points out that the second step in the dolomite growth is rate controlling, which means that the retained water molecules adhered to surface $Mg^{2+}$ block the $CO_3^{2-}$ ions in the solution from being sufficiently attracted by surface $Mg^{2+}$ ions.[4] Therefore, the key role of dissolved sulfide in promoting dolomite formation is very likely to lie in the dehydration of $Mg^{2+}$ ions at the (proto)dolomite or other precursor surface. However, the effect of hydrogen sulfide on this process has never been explored. A hint may come from a related work on calcite. Specifically, atomic force microscopy (AFM) work and molecular dynamic (MD) simulations have shown that ethanol and polysaccharides can be more strongly bound to the calcite surface than water, thus repelling the water molecules from the surface and forming a hydrophobic layer.[21-23] Meanwhile, ethanol and polysaccharides have been proven to enhance $Mg^{2+}$ incorporation into precipitating Ca-Mg carbonate.[15,24] For the dolomite case, one explanation is that the surface water molecules can be removed by adsorbed hydrogen sulfides on the dolomite surface. In order to study the thermodynamics of the adsorption processes, we performed quantum-mechanical calculations based on the density functional theory (DFT).

## METHODS

### Computational details

We use DFT as implemented in the Vienna *ab initio* simulation package (VASP).[25] The general gradient approximation (GGA) with the Perdew, Burke, and Ernzerhof (PBE) parameters is employed.[26] The projector-augmented wave (PAW) method with the energy cutoff of 600 eV has been used. All structures are relaxed using both the static energy minimization scheme and *ab initio* molecular dynamics simulations at 10K. First, we determine energies of adsorption of a monolayer of either water or $H_2S$ onto the dolomite {104} surface, which is the main cleavage and growth plane of dolomite. Both vacuum and solution are considered as reference states for adsorption calculations. The pH of modern seawater and pore water in modern dolomite sediments is usually 7~8.2[8, 27-29] with some exceptions in modern lacustrine environments where pH of lake water and pore water are larger than 9[29]. There are two major species of dissolved sulfide: $HS^-$ and $H_2S$. When pH is between 7 and 8, aqueous $H_2S$ accounts for 10~50% of total dissolved sulfides.[30] When pH is greater than 8, the surface charge of dolomite is negative at $pCO_2 = 10^{-3.5}$ atm.[31] As a result, it is most likely that it is $H_2S$ that is concentrated on the dolomite surface at this pH condition, although $HS^-$ is the dominant solution species. To consider the two pH ranges, calculations were performed for H₂S instead of HS⁻. Dolomite {104} surfaces have been simulated using 4 layer slabs of a triclinic unit cell (Figure 2) that has the following parameters: $a$=9.085Å; $b$=4.812Å; $c$=4.812Å; $\alpha$=120.00°; $\beta$=37.40°; $\gamma$=120.00°. The top and the bottom free surfaces have identical structures and periodic boundary conditions are used in all spatial directions. For a monolayer of either water or H₂S adsorption on the surface, a vacuum of ~14 Å thickness has been inserted above one of the surfaces in order to prevent the interactions between the bottom surface and the water or H₂S molecules adsorbed to the top surface. The bottom layer of the dolomite surface (10 atoms) is fixed for the calculations of adsorption in vacuum. Calculations for cases where both surfaces have an adsorbed monolayer were also performed and the adsorption energies per surface unit cell are the same as one surface adsorption case. To simulate the adsorption in a solution environment, a bulk water space of 14Å with density of ~1 g/cm³ was inserted above the surface with an adsorbed monolayer. All layers of the dolomite slab are allowed to relax in solution calculations. To ensure a minimum energy configuration of the water surrounding the dolomite, we took multiple samples of water structures generated from classical MD simulation using the code LAMMPS[32] and TIP4P potential[33]. Two small volumes of water corresponding to 20 water molecules were taken from the previous structures. We ensured the densities of the two water structures and relaxed them again in DFT. In order to account for the weak van der Waals interactions in the adsorptions systems, for adsorptions in solution we performed calculations using the DFT-D2 method of Grimme[34]. We tested k-point convergence with a criterion of 1 meV/atom and a mesh of 3×3×1 was used for both vacuum and solution systems.

**Thermodynamic model**

To test whether aqueous $H_2S$ molecules are more strongly adsorbed to the dolomite {104} surface than water we compare the energies of the initial and final steps in the following reaction. Initially, a monolayer of water molecules with 100% coverage (one water molecule above each cation) is adsorbed on the dolomite {104} surface with the $H_2S$ molecules far away from the surface in the solution. To present the end of the reaction, a monolayer of $H_2S$ molecules with 100% coverage was put on the surface replacing the surface water molecules. The free energy of this reaction at room temperature can be expressed as

$$\Delta G_{ads} = [\Delta G_{dol+H_2S}(T) - \Delta G_{dol+H_2O}(T)] - [\mu_{H_2S}(T) - \mu_{H_2O}(T)], \quad (1)$$

where $\Delta G_{dol+H_2S}(T)$ and $\Delta G_{dol+H_2O}(T)$, respectively, are the Helmholtz free energy of $H_2S$ and $H_2O$ monolayers adsorbed to dolomite at room temperature (either in vacuum or in solution as described in the previous section); $\mu_{H_2S}(T)$ and $\mu_{H_2O}(T)$ are chemical potentials of an aqueous $H_2S$ molecule and a liquid water molecule at room temperature, respectively. If $\Delta G_{ads} < 0$, then adsorption of $H_2S$ is more energetically favorable than adsorption of water and vice versa. The energy of adsorption system at room temperature for any of our systems is composed of three parts: the DFT energy calculated at 0K, the zero point energy (ZPE), and the thermal energy of vibration. For ZPE, only the vibrational energies of the surface adsorbed atoms are relevant for our study. Our tests show that the major contributions to the vibrational energy are from the molecules of the first four layers above the surface. The free energy of each adsorption system can be expressed as:

$$\Delta G_{dol+H_2S}(T) = E_{dol+adsb,vasp} + \sum_i \frac{1}{2} h\nu_i + RT \sum_i \ln(1 - e^{-h\nu_i/k_B T}) \quad (2)$$

The first two terms on the right hand side of Eq (1) can be approximated as:

$$\Delta G_{dol+H_2S} - \Delta G_{dol+H_2O} = E_{dol+H_2S,vasp} - E_{dol+H_2O,vasp} + (G_{surf+H_2S} - G_{surf+H_2O}) \quad (3)$$

Chemical potentials of a liquid water molecule and an aqueous $H_2S$ at room temperature can be calculated as

$$\mu_{H_2O}^{liquid}(T) = E_{H_2O}^{molecule} + \Delta G_{H_2O}^{excitations}(T) - \Delta G_{H_2O}^{vaporization}(T)$$

$$= E_{H_2O}^{molecule} - 1.06 \; kJ/mol \tag{4}$$

$$\mu_{H_2S}^{aq}(T) = E_{H_2S}^{molecule} + \Delta G_{H_2S}^{excitations}(T) - \Delta G_{H_2S}^{solvation}(T)$$

$$= E_{H_2S}^{molecule} - 5.76 \; kJ/mol \tag{5}$$

The general equation for excitation energy is as below[34-35]:

$$\Delta G^{excitations}(T) = -RT \ln \left\{ \left[ \frac{2\pi (\sum_i m_i) k_B T}{h^2} \right]^{\frac{3}{2}} \frac{V_e}{N} \right\} - RT \ln \left[ \frac{\pi^{\frac{1}{2}}}{\sigma} \left( \frac{T^3}{\theta_A \theta_B \theta_C} \right)^{\frac{1}{2}} \right] + RT \sum_i^3 \left[ \frac{h\nu_i}{2k_B T} + \ln\left(1 - e^{-h\nu_i/k_B T}\right) \right] \tag{6}$$

where σ is symmetry number (=2 for water and $H_2S$ molecules), θ is rotational temperature and $\nu_i$ is vibrational frequency with the *i*th normal mode. The three terms at the right hand side describe the translational, rotational and vibrational contributions to excitation energy respectively. ZPE is included in the third term. The parameters for a water molecule have been described in McQuarrie[35,36], and the parameters for $H_2S$ in Senekowitsch and co-workers[37] and Hoffman and co-workers[38]. The vaporization energy and solvation energy are taken from the difference of free energy between gaseous water and liquid water and between gaseous $H_2S$ and aqueous $H_2S$ at 298.15K, 1 atm, standard states.[39] The energy of a single $H_2O$ and a $H_2S$ molecule at 0 K were calculated in VASP in a 10×10×10 Å supercell and a 15×15×15 Å supercell, respectively. By combining Eqs. (1) to (4), we obtain

$$\Delta G_{ads} = \{(E_{dol+H_2S,vasp} - E_{dol+H_2O,vasp}) + (G_{surf+H_2S} - G_{surf+H_2O})\} - \{(E_{H_2S}^{molecule} - E_{H_2O}^{molecule}) + (\Delta G_{H_2S}^{excitations} - \Delta G_{H_2O}^{excitations}) + (\Delta G_{H_2S}^{solvation} - \Delta G_{H_2O}^{vaporization})\} \quad (7)$$

The adsorption energy of a mixed layer of H₂O and H₂S from solution can expressed as

$$\Delta G = E_b^{H_2S} \cdot n_{H_2S} + E_b^{H_2O} \cdot n_{H_2O} - TS_{config}(n_{H_2S}, n_{H_2O}) \quad (8)$$

where $E_b^{H_2S}$ and $E_b^{H_2O}$ are adsorption energies of a H₂O and a H₂S molecule, respectively, and the two adsorbed species are not interacting. $n_{H_2S}$ and $n_{H_2O}$ are the numbers of adsorbed molecules of H₂S and H₂O, respectively. These adsorption energies can in turn be written as

$$E_b^{H_2O} = U_{dol-H_2O} - \mu_{bw} n_{bw} - E_{dol} \quad (9)$$

and

$$E_b^{H_2S} = U_{dol-H_2S} - \mu_{H_2S} n'_{H_2S} - \mu_{bw} n'_{bw} - E_{dol} - E_b^{H_2O} \quad (10)$$

where $\mu_{bw}$ is the chemical potential of bulk water, and $E_{dol}$ is the energy of the dolomite slab. In the above expression,

$$\mu_{H_2S} = E_{H_2S}^{VASP} - 5.76 \frac{kJ}{mol} + RT\ln(X_{H_2S}^{Sol}) \quad (11)$$

where $X_{H_2S}^{Sol}$ is the concentration of H₂S in solution. The configurational entropy in Eq. (8) can be calculated as

$$S_{config} = -R[n_{H_2S}\ln(X_{H_2S}) + n_{H_2O}\ln(X_{H_2O})] \quad (12)$$

In order to minimize the energy, the first order derivatives of ΔG are equated to zero as follows

$$\frac{\partial \Delta G}{\partial n_{H_2S}} = \frac{\partial}{\partial n_{H_2S}}(n_{H_2S} E_b^{H_2S} + (n_T - n_{H_2S}) E_b^{H_2O} - TS_{config})$$

$$= E_b^{H_2S} - E_b^{H_2O} + RT\ln(n_{H_2S}/(n_{H_2O} \cdot X_{H_2S}^{Sol})) = 0 \quad (13)$$

where $n_T$ is the total number of adsorption sites, respectively. $n_T$ is estimated to be 14 μmol/m² for 100% coverage[31], which corresponds to one water molecule above each cation.

**RESULTS**

**Adsorption in vacuum**

Six configurations with different $H_2O$ orientations were first explored by *ab initio* MD simulations and then optimized by static relaxations. In each configuration, a water molecule lying parallel to the dolomite (104) surface was initially placed above each cation of the surface (total of 2 water molecules per surface unit cell). The initial distance between the O atom of water and the surface cation was chosen to be 2.4Å, which is comparable to the Ca-$O_w$ (O of water) distance for calcite[40,41]. In the lowest energy configuration, orientations of water molecules are different above surface $Mg^{2+}$ and $Ca^{2+}$ ions. Specifically, the water molecule above Mg relaxes so that the Mg-$O_w$ distance is 2.17Å, one hydrogen binds to an O atom of the surface carbonate group by hydrogen bond (with H-$O_{carb}$ distance of 1.83Å) and another H points away from the surface (Figure 3). Above Ca, the water molecule assumes a position such that the Ca-$O_w$ distance is 2.41Å and both hydrogen atoms bind to surface $O_{carb}$ (one is strong with H-$O_{carb}$ distance of 1.75Å and another one weak with H-$O_{carb}$ distance of 2.11Å). Compared to an earlier classical MD simulation result, which showed the average Mg-$O_w$ distance of 3.02Å and the average Ca-$O_w$ distance of 2.45Å[42], our *ab initio* results not only show stronger surface cation-water binding (manifested in shorter bonds), but also reverse the previous conclusion that $Ca^{2+}$ ions are more strongly bound to water than $Mg^{2+}$ ions. Six similar initial configurations where $H_2O$ was replaced by $H_2S$ were also considered in our calculations. In the lowest energy configuration, the $H_2S$ molecules display similar trends in bond lengths above $Mg^{2+}$ and $Ca^{2+}$ ions to the one found for $H_2O$ (see Table 1). The main difference between $H_2S$ and $H_2O$ is that in the case of $H_2S$ only one hydrogen is bound to the surface $O_{carb}$ with H-$O_{carb}$ distances of 2.02Å and 1.86Å for $Mg^{2+}$ and $Ca^{2+}$ sites, respectively (Figure 3). The adsorption energy of a water molecule on dolomite surface has shown to be 68.07 kJ/mol in a recent DFT calculation.[43] By using the same equation as $E_{ads}= E^{surf+adsorbate} - E^{surf} - E^{adsorbate}$, we obtain a similar result: 71.16 kJ/mol. The adsorption energy of a water molecule on dolomite (104) surface at 0K is 34.5kJ/mol lower than $H_2S$ (Table 2).The energy difference narrows to 13.6 kJ/mol with ZPE and entropy corrections, but it does not change the trend that water is more stable on the dolomite surface.

**Adsorption in solution**

The bulk water was initially placed in various positions above the monolayer of water molecules adsorbed on dolomite (104) surface in such a way that the average distance between the oxygen atoms in the bottom layer of bulk water and the highest oxygen atoms in surface carbonate groups ranged from 0.7Å to 3.7Å in increments of 0.5 Å. Each step was calculated by both *ab initio* MD and static relaxation and the lowest energy configuration from this test was found to have the bulk water-surface distance of 3.18Å. In order to further explore the effect of surface water density, we performed additional *ab initio* MD calculations with one extra water molecule inserted between the bulk water and the adsorbed first layer. Five lowest energy configurations from the *ab initio* MD calculations were then optimized in static relaxation. The energies of the five optimized configurations are very similar (table 1, the difference is less than 1 meV/atom) and are lower than that of the configuration with lower water density. The energies reported in the table 2 are the average value of the 5 configurations. A series of similar calculations have been carried out on the $H_2S$ adsorption system. The configurations with the lowest energies (table 1) were also obtained by inserting an extra water molecule. In further analysis and discussion, we focus on the structures with a higher water density, since they were shown above to have lower energies. With the presence of bulk water, the adsorption energy difference between a $H_2O$ and a $H_2S$ molecule is 41.9 kJ/mol (at 0K) and 12.8 kJ/mol (at room temperature) (Table 2). ZPE is the major contribution to the corrections. Based on the geometry of surface molecules (Figure 4) and a series ZPE calculations of different surface layers (the ΔZPE ($= ZPE_{H_2S}^{surf} - ZPE_{H_2O}^{surf}$) of the atoms in the first 5 layers is only 0.1 kJ/mol higher than those in the first 4 layers), the ZPEs of the adsorbed layer and the first three bulk water layers (6 water molecules) are included in our result. The calculation including bulk water still shows the favorable adsorption of water over $H_2S$.

**The adsorption of a mixed layer**

Although water is more energetically stable on dolomite surface than aqueous hydrogen sulfide, our two-phase adsorption model predicts a small amount of $H_2S$ adsorbed on the surface depending on the adsorption energy difference and pH of solution (affects the concentration of aqueous $H_2S$). At pH of 7.0~8.2 where the modern dolomite is precipitating, 6%~50% of dissolved sulfide are in aqueous $H_2S$ phase. Usually more than 5mM dissolved sulfide was measured in pore water from modern dolomite site.[44] According to equation 13, the density of aqueous $H_2S$ on the surface can be up to ~2000 molecule/$\mu m^2$ at this pH condition and 5 mM dissolved sulfide concentration (the energy difference of 12.8 kJ/mol was used). In order to test the effect of the surface $H_2S$ on the $Mg^{2+}$-water complex, a mixed layer of 50% $H_2S$ and 50% $H_2O$ molecules was placed above the dolomite surface in the presence of bulk water. We tested multiple configurations by using the same method as described in previous section and the lowest energy structure was obtained by inserting an extra water molecule between the bulk water and the adsorbed first layer. In the adsorbed mixed layer, the $Mg^{2+}$-$O_w$ distance increases from ~2.18 Å to ~2.23 Å compared to the respective case of pure layers (Table 1).

**DISCUSSION**

Our calculations show that in vacuum, water is more stable than aqueous $H_2S$ on the dolomite (104) surface at room temperature with the difference in adsorption energy of -13.6 kJ/mol per adsorbed molecule. The added bulk water does not change the trend that it is thermodynamically favorable for water adsorption with the energy gain of -12.8 kJ/mol per adsorbed molecule. In both cases, ZPE and other corrections narrow the energy difference between water and $H_2S$ adsorptions but do not change the conclusion. In summary, calculations under both vacuum and solution conditions do not support the explanation that aqueous $H_2S$ molecules, preferentially adsorbed over $H_2O$, form a hydrophobic layer on dolomite surface.

The multiphase adsorption model predicts that ~0.002 molecule/nm² of aqueous $H_2S$ can be adsorbed at pH condition (7~8.2) and dissolved sulfide concentration (~5 mM) close to some modern environments. This concentration corresponds to a little less than 1 ‰ of surface sites, At some local environments where dolomite and high magnesium calcite precipitate, the concentration of dissolved sulfide can be even much higher[45-48] (up to 20mM). Thus, the effect of adsorbed $H_2S$ on the surface magnesium hydration bond is also important to the understanding of the role of dissolved hydrogen sulfide. Generally for heterogeneous catalysis, a good catalyst needs to bind to the reactant strongly enough but not too strongly. Similarly, the strong surface $Mg^{2+}$-water bond inhibits the dolomite growth, while the relatively weaker surface-$H_2S$ interaction likely increases the competence of $CO_3^{2-}$ to bond to the $Mg^{2+}$.

On the other hand, according to our study, the adsorbed aqueous $H_2S$ on the surface can affect the $Mg^{2+}$-$H_2O$ bond distance. Because of the relatively larger size of a $H_2S$ molecule, the $H_2S$ molecules need extra space on the surface, which creates a local environment for relieving surface water molecules from being constrained by the surrounding bulk. When the constraint is released to a certain degree by the local environment created by large $H_2S$ molecules, the surface water molecules relax close to the positions they have without constraint. Therefore, another possibility for $H_2S$ facilitating the carbonation is that, similarly to the $H_2S$ effect on $Mg^{2+}$-$H_2O$ bond distance, there is room for direct interaction between $Mg^{2+}$ and $CO_3^{2-}$ due to the geometry and the large size of $H_2S$ and larger space between $H_2O$/$H_2S$ and dolomite surface. Detailed mechanisms for $CO_3^{2-}$ adsorption in the presence of $H_2S$ will be explored in our future studies.

ACKNOWLDEGEMENTS

Zhizhang Shen and Huifang Xu acknowledge funding from NSF (EAR-095800) and NASA Astrobiology Institute (N07-5489) . Izabela Szlufarska and Yun Liu acknowledge NSF EAR-0910779 grant.

**Figure 1.** The proposed processes for the growth of dolomite crystal catalyzed by dissolved hydrogen sulfide. First, partially dehydrated $Mg^{2+}$ ions are adsorbed onto the dolomite surface. The remaining water bonds to surface $Mg^{2+}$ ions and blocks the approach of carbonate group to the $Mg^{2+}$ ions which is enabled by the presence of $H_2S$.

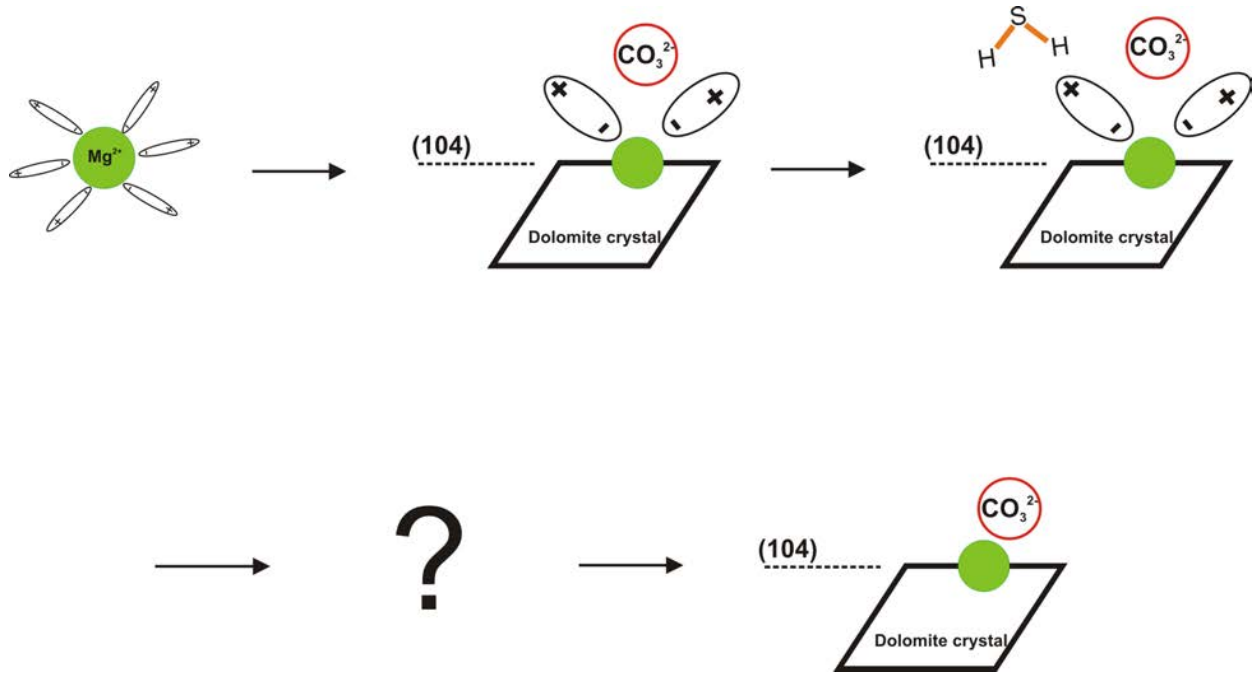

**Figure 2.** A four-layer slab model of dolomite with (104) surfaces were used in this study (view of 2×2×1 supercell). The O, Mg, Ca and C atoms are displayed as red, orange, cyan, and brown, respectively.

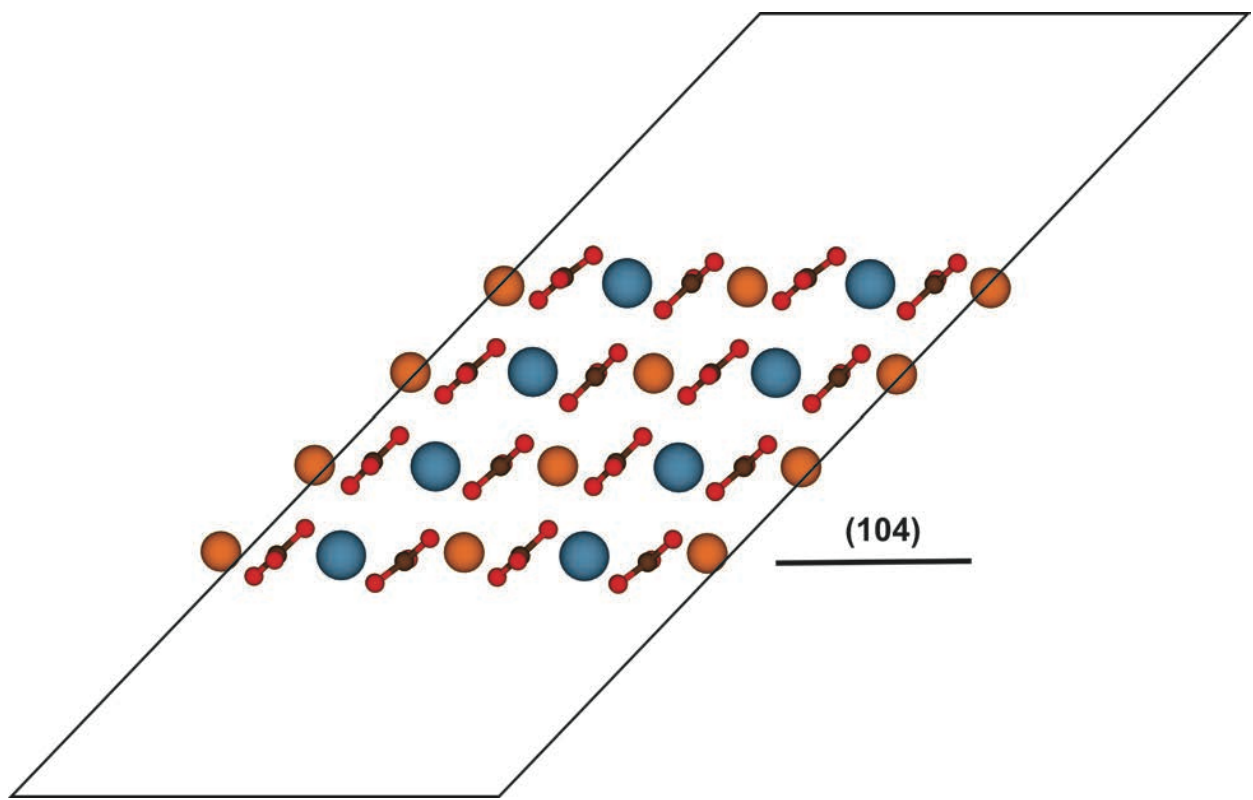

**Figure 3.** Left: 1 monolayer of water molecules adsorbed on dolomite (104) surface. Right: 1 monolayer of H₂S molecules adsorbed on dolomite (104) surface. Red: O, yellow: S, orange: Mg, cyan: Ca, brown: C, and white: H.

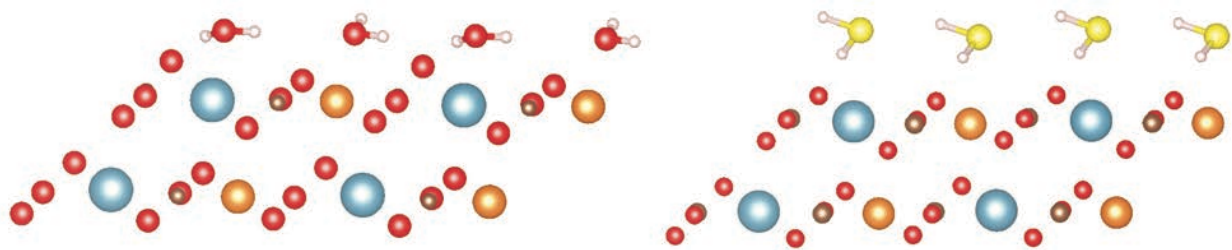

**Figure 4.** Top: 1 monolayer of water molecules in solution adsorbed on dolomite (104) surface. Below: 1 monolayer of H₂S molecules adsorbed in solution on dolomite (104) surface. Red: O, yellow: S, orange: Mg, cyan: Ca, brown: C, and white: H. Only atoms of the first layer of the surface are shown.

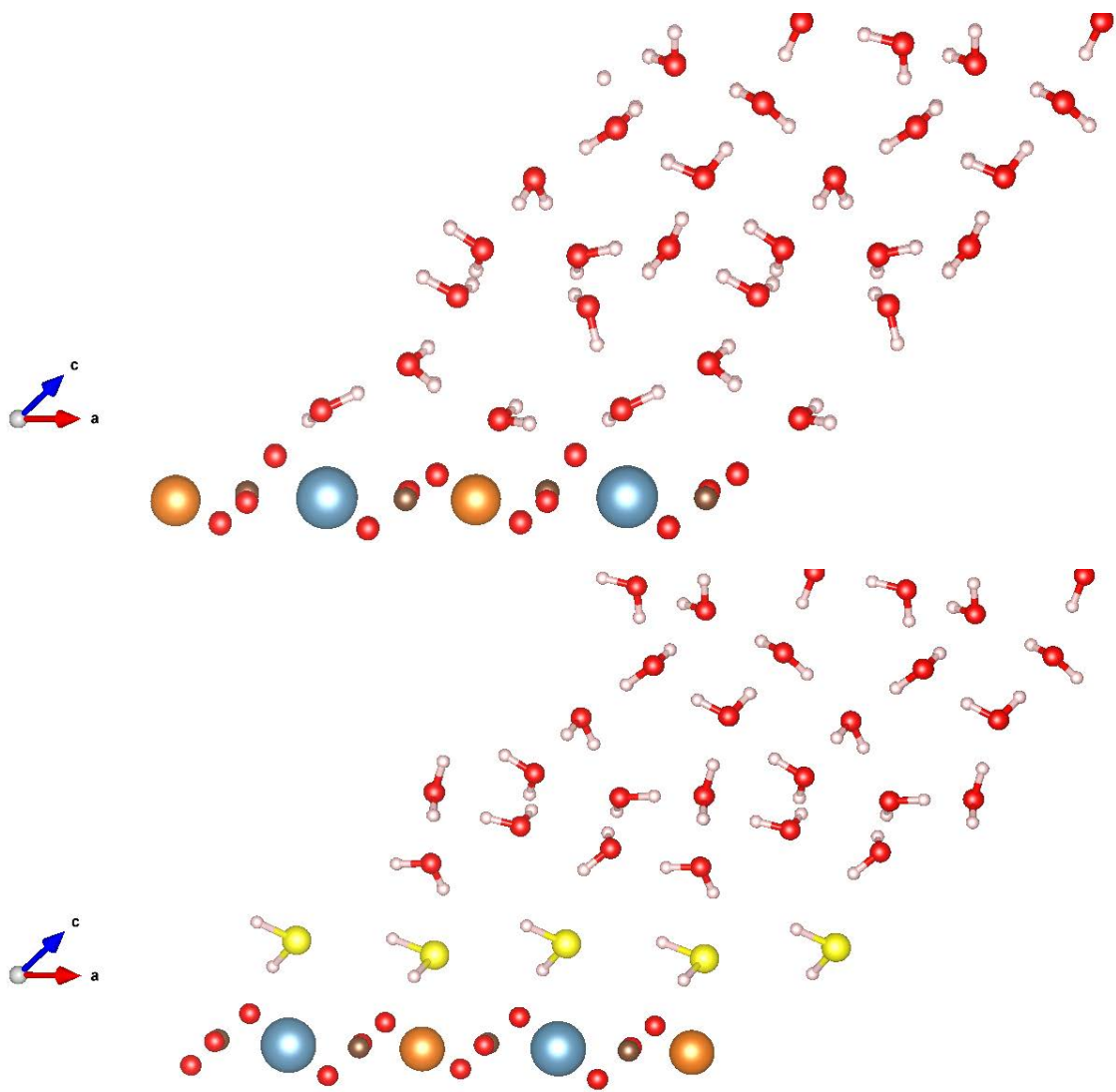

**Table 1.** Energies and distances between surface Mg and O of adsorbed water for adsorption of a) 1 mono-layer (ML) of $H_2O$ in solution; b) 1 ML of $H_2S$ in solution; and c) a mixed $H_2O$ and $H_2S$ layer in solution.

| Adsorbate | $H_2O$ | | $H_2S$ | Mixture | |
|---|---|---|---|---|---|
| Configuration | Energy (eV) | Mg-O (Å) | Energy (eV) | Energy (eV) | Mg-O (Å) |
| 1 | -301.436 | 2.184 | -297.934 | -299.862 | 2.237 |
| 2 | -301.436 | 2.176 | -297.925 | -299.861 | 2.231 |
| 3 | -301.435 | 2.176 | -297.926 | -299.862 | 2.224 |
| 4 | -301.436 | 2.185 | -297.931 | -299.859 | 2.227 |
| 5 | -301.432 | 2.174 | -297.923 | -299.863 | 2.229 |

**Table 2.** Energy difference between the adsorption of $H_2S$ and $H_2O$ on dolomite (104) surface. The energies shown in this table are in kJ/mol.

|  | In vacuum | | In solution | |
|---|---|---|---|---|
| $\Delta H_{abinitio}$ | 331.1 | 331.1 | 338.5 | 338.5 |
| $\Delta G_{vib}$ |  | -25.6 |  | -33.8 |
| $\Delta E_{H2O-H2S}$ | -296.6 | -296.6 | -296.6 | -296.6 |
| $\Delta G_{excitation}$ |  | 19.0 |  | 19.0 |
| $\Delta G_{solv-vapor}$ |  | -14.3 |  | -14.3 |
| **Net energy** | 34.5 | 13.6 | 41.9 | 12.8 |

SUPPLEMENTAL MATERIAL

The derivations of chemical potentials of water and aqueous H₂S are shown below:

(1) The expression for the water chemical potential at finite temperature (Pinney et al., 2009):
$$\Delta\mu_{H_2O}^{liquid}(T) = E_{H_2O}^{molecule} + \Delta G_{H_2O}^{exciations}(T) - \Delta G_{H_2O}^{vaporization}(T)$$
where T=298.15K.
The excitation energy can be expressed as: $\Delta G_{H_2O}^{exciations}(T) =$

$$-RT \ln\left\{\left[\frac{2\pi(\Sigma_i m_i)k_B T}{h^2}\right]^{\frac{3}{2}} \frac{V_e}{N}\right\} - RT \ln\left[\frac{\pi^{\frac{1}{2}}}{\sigma}\left(\frac{T^3}{\theta_A \theta_B \theta_C}\right)^{\frac{1}{2}}\right] + RT \sum_i^3 \left[\frac{h\nu_i}{2k_B T} + \ln(1 - e^{-h\nu_i/k_B T})\right]$$

where σ is symmetry number (=2 for water and H₂S molecules), θ is rotational temperature and $\nu_i$ is vibrational frequency with the *i*th normal mode. The three terms at the right hand side describe the translational, rotational and vibrational contributions to excitation energy respectively.

$\Theta_{rot}$ = 40.1, 20.9, 13.4 K,  σ =2 (symmetry number)
$\Theta_{vib,i}$ = 5360, 5160, 2290 K
The parameters above are obtained from McQuarrie and Simon, 1999.
Debye temperatures are related to vibrational frequencies through this equation: $\theta_D = h\nu/k_B$. Put all the parameters into the equation above, we can get:
$G_{trans}$ = -37.00 $kJ/mol$
$G_{rot}$ = -9.33 $kJ/mol$
$G_{vib}$ = 53.88 $kJ/mol$
Combine these three numbers, we can get: $\Delta G_{H_2O}^{exciations}(298.15K) = 7.55\ kJ/mol$
$\Delta G_{H_2O}^{vaporization}(298.15K) = -8.61\ kJ/mol$
The vaporization data are from Wagman et al., 1968.
So,
$$\Delta H_{H_2O}^{liquid}(T) = E_{H_2O}^{molecule} - 1.06\ kJ/mol$$

(2) For aqueous H₂S chemical potential at finite temperature

Similarly,
$$\Delta\mu_{H_2S}^{aq}(T) = E_{H_2S}^{molecule} + \Delta G_{H_2S}^{exciations}(T) - \Delta G_{H_2S}^{solvation}(T)$$

The vibrational frequencies (cm⁻¹) of a H₂S molecule (Hoffmann et al., 1997):

ν1= 2721.05    symmetry stretching

v2= 1214.0     bend

v3= 2729.3     asymmetry stretching

$\Theta_{rot}$ = 14.9, 12.93, 6.93 K (Senekowitsch et al., 1988)
$G_{trans}$ = -39.35 $kJ/mol$
$G_{rot}$ = -11.95 $kJ/mol$
$G_{vib}$ = 39.85 $kJ/mol$
$\Delta G_{H_2S}^{exciations}(298.15K) = -11.45 kJ/mol$

At 298.15K, 1 atm, standard state (Wagman et al., 1968):

H₂S (g): $\Delta G_f^0 = -33.56 \, kJ/mol$

H₂S (aq): $\Delta G_f^0 = -27.87 \, kJ/mol$

$\Delta G_{H_2O}^{solvation}(298.15K) = 5.69 \, kJ/mol$

In sum,

$\Delta \mu_{H_2S}^{aq}(T) = E_{H_2S}^{molecule} - 5.76 \, kJ/mol$

References for supplemental material